\documentclass[%
 reprint,
 superscriptaddress,
 amsmath,amssymb,
 aps,
 prb,
 floatfix,
]{revtex4-2}

\usepackage{graphicx}
\usepackage{dcolumn}
\usepackage{bm}

\begin{document}
\title{
Dimensional change of quadrupole orders in pseudospin-$\frac{1}{2}$ pyrochlore magnets under [111] field
}

\author{Hiroaki Kadowaki}
\affiliation{Department of Physics, Tokyo Metropolitan University, Hachioji, Tokyo 192-0397, Japan}

\author{Hiroshi Takatsu}
\affiliation{Department of Energy and Hydrocarbon Chemistry, Graduate School of Engineering, Kyoto University, Kyoto 615-8510, Japan}

\author{Mika Wakita}
\affiliation{Department of Physics, Tokyo Metropolitan University, Hachioji, Tokyo 192-0397, Japan}

\date{\today}

\begin{abstract}
We have studied long range orders of electric quadrupole moments 
described by an effective pseudospin-$\frac{1}{2}$ Hamiltonian representing 
pyrochlore magnets with non-Kramers ions under [111] magnetic field, 
in relevance to Tb$_{2}$Ti$_{2}$O$_{7}$. 
Order parameters and phase transitions of this frustrated system 
are investigated using classical Monte-Carlo simulations. 
In zero field, the model undergoes a first-order phase transition 
from a paramagnetic state to an ordered state 
with an antiparallel arrangement of pseudospins. 
This pseudospin order is characterized by the wavevector $\bm{k}=0$ 
and is selected by an energetic or an order-by-disorder mechanism from 
degenerate $\bm{k}=(h,h,h)$ mean-field orders. 
Under [111] magnetic field this three-dimensional quadrupole order 
is transformed to a quasi two-dimensional quadrupole order 
on each kagom\'{e} lattice separated by field-induced 
ferromagnetic triangular lattices. 
We discuss implication of the simulation results with respect to experimental 
data of Tb$_{2}$Ti$_{2}$O$_{7}$.
\end{abstract}

\maketitle

\section{Introduction}
Magnetic systems with geometric frustration have been studied 
experimentally and theoretically for decades \cite{Lacroix11}. 
In particular, systems on networks of triangles or tetrahedra, 
such as triangular \cite{Wannier50}, kagom\'{e} \cite{Shyozi51,Qi2008}, 
and pyrochlore \cite{Gardner10} lattices, show interesting behavior 
due to the frustration. 
Among them classical spin ice on the pyrochlore lattice \cite{Bramwell01} 
has been investigated in depth 
from viewpoints of the finite zero-point entropy of water ice \cite{Ramirez99}, 
field-induced two-dimensional (2D) kagom\'{e} ice \cite{MatsuhiraJPCM2002,Tabata06,Fennell2007},
emergent magnetic monopoles \cite{Castelnovo08,Kadowaki09},
topological sectors \cite{Jaubert2013}, etc.  
In recent years quantum spin liquid (QSL) states \cite{Lee08,Balents10}, 
where conventional long-range orders (LRO) are suppressed 
by quantum fluctuations, are being intensively studied \cite{Savary2017}. 
A QSL state is theoretically predicted 
for spin-ice like systems \cite{Hermele04,Benton12,Lee12,Gingras14}, 
where transverse spin interactions transform 
the classical spin ice into QSL. 

Among frustrated magnetic pyrochlore oxides \cite{Gardner10} 
Tb$_{2+x}$Ti$_{2-x}$O$_{7+y}$ (TTO) has attracted much attention 
as a QSL candidate, because 
no conventional magnetic orders have been found \cite{Gardner99,Kadowaki2018}, 
and a quantum version of spin ice was theoretically proposed \cite{Molavian07,Gingras14}. 
Recently we showed that the putative QSL state of TTO 
is limited in a range of the small off-stoichiometry parameter 
$x < x_{\text{c}} \simeq -0.0025$ \cite{Taniguchi13,Wakita16,Kadowaki2018}. 
While in the other range $x_{\text{c}} < x$ TTO undergoes a phase transition 
most likely to an electric multipolar (or quadrupolar) state 
($T<T_{\text{c}}$) \cite{Takatsu2016prl} 
which is described by an effective pseudospin-$\frac{1}{2}$ 
Hamiltonian for non-Kramers ions \cite{Onoda11}. 
The estimated parameter set of this Hamiltonian \cite{Takatsu2016prl} 
is close to the theoretical phase boundary 
between the electric quadrupolar state and a U(1) QSL state \cite{Onoda11,Lee12},
which is hence a theoretical QSL candidate for TTO. 

In our previous investigations 
using a TTO crystal sample with $T_{\text c}=0.53$ K \cite{Takatsu2016prl,Takatsu16,Takatsu2017}, 
specific heat and magnetization 
under [111] and [100] magnetic fields 
were measured and 
finite-temperature phase-transitions 
were semi-quantitatively analyzed using 
classical Monte-Carlo (CMC) simulation techniques. 
Despite the quantum nature of the pseudospin-$\frac{1}{2}$ Hamiltonian \cite{Onoda11,Lee12}, 
the classical treatment provided us good arguments 
that TTO can be described by the Hamiltonian \cite{Takatsu2016prl}. 
Although quantum (e.g. \cite{Bojesen2017}) and classical (e.g. \cite{Zhitomirsky14}) 
properties of these types of pseudospin-$\frac{1}{2}$ Hamiltonians 
for non-Kramers and Kramers pyrochlore magnets are of interest, 
they have not been fully investigated \cite{Rau_Gingras2018}. 

In this paper we present detailed studies of CMC simulations 
to complement our previous study of the quadrupole orders in TTO \cite{Takatsu2016prl}. 
In particular, order parameters and finite-temperature phase transitions 
of the quadrupolar states were remained to be elucidated 
from a theoretical standpoint \cite{Takatsu2016prl}.
We have shown that under zero and low [111] fields 
the quadrupole ordered states have three dimensional (3D) and 2D 
characters, respectively. 
Nature of these phase transitions in zero and low fields 
is shown to be first order and second order 
with the 2D Ising universality class, respectively. 
Implication of the CMC simulation results is discussed 
with respect to experimental data of TTO.

\section{Effective pseudospin-$\frac{1}{2}$ Hamiltonian and CMC simulation}
The minimal pseudospin-$\frac{1}{2}$ Hamiltonian 
for TTO \cite{Takatsu2016prl,Kadowaki15} is described by 
\begin{align}
\mathcal{H} = 
J_{\text{nn,eff}} \sum_{\langle {\bm r} , {\bm r}^{\prime} \rangle} 
& \sigma_{\bm{r}}^{z} \sigma_{\bm{r}^{\prime}}^{z} 
- J_{\text{nn,eff}} \bm{H} \cdot \sum_{\bm r} \bm{z}_{\bm{r}} \sigma_{\bm{r}}^{z} \nonumber \\
+ J_{\text{nn,eff}} \sum_{\langle {\bm r} , {\bm r}^{\prime} \rangle} 
&[2 \delta ( \sigma_{\bm{r}}^+ \sigma_{\bm{r}^{\prime}}^- 
           + \sigma_{\bm{r}}^- \sigma_{\bm{r}^{\prime}}^+ )  \nonumber \\
&+ 2 q ( e^{2 i \phi_{\bm{r},\bm{r}^{\prime}} } 
\sigma_{\bm{r}}^+ \sigma_{\bm{r}^{\prime}}^+ + \text{H.c.} ) ] \; , 
\label{H_effective}
\end{align}
where the first and second terms are magnetic interactions: 
nearest-neighbor (NN) superexchange interaction of magnetic moment operators 
$\sigma_{\bm{r}}^{z}$ (the Pauli matrix) acting on the crystal field (CF) 
ground state doublet at a site $\bm{r}$, 
and Zeeman energy under dimensionless external magnetic field $\bm{H}$. 
These magnetic terms has been used as the model of spin ice 
with the effective coupling constant $J_{\text{nn,eff}}$ ($>0$) \cite{Hertog00}. 
The third term of Eq.~(\ref{H_effective}) 
represents NN superexchange interaction of quadrupole moment operators 
$\sigma_{\bm{r}}^{\pm}= ( \sigma_{\bm{r}}^x \pm i \sigma_{\bm{r}}^y )/2$ \cite{Onoda11}. 
This term induces quantum fluctuations to the classical spin ice for 
the non-zero dimensionless parameters $\delta$ and $q$. 
Other detailed definitions of Eq.~(\ref{H_effective}), 
the lattice site, its local axes etc. \cite{Kadowaki15,Takatsu2016prl}, 
are described in the appendix. 

In Eq.~(\ref{H_effective}) we omit the dipolar interaction included in Eq.~(1) of 
Ref.~\cite{Takatsu2016prl} in order to perform CMC simulations with larger system sizes. 
In this simplification, the typical parameters of the Hamiltonian for TTO are 
$J_{\text{nn,eff}}=1.48$ K, $\delta=0$, and $q = 0.57$ \cite{ParametersJdeltaq}.
In zero field, the classical ground state of Eq.~(\ref{H_effective}) with these parameters 
is LRO of $xy$-components of the pseudospins (quadrupole order), 
which is denoted by the planar antiferropseudospin (PAF) phase 
(Fig.~7 in Ref.~\cite{Onoda11}). 

By treating the pseudospin $\bm{\sigma}_{\bm{r}}$ as a classical unit vector \cite{LandauBinder15}, 
we carried out CMC simulations of the classical spin model described by Eq.~(\ref{H_effective}).
Since critical behaviors of finite-temperature phase-transitions 
are expected to be the same for classical and quantum models \cite{Sachdev11,Zhitomirsky14}, 
CMC simulations can be used to shed light on experimental data. 
For present CMC simulations we used parameter sets 
in a range relevant to TTO: $-0.1 \le \delta \le 0.1$ 
and $0.2 \le q \le 0.7$ \cite{Takatsu2016prl},  
which encompasses the PAF and classical spin ice states \cite{Onoda11}. 
These simulations were performed typically 
with $\sim 4 \times 10^6$ MC steps per spin 
and for periodic clusters with $N= 12 L \times L \times L^{\prime} \leq 629856$ spins, 
where $L$ and $L^{\prime}$ stand for linear dimensions 
perpendicular and parallel to a [111] direction, respectively. 
The magnetic field was applied parallel to this [111] direction, 
along which there are $3 L^{\prime}$ triangular layers and $3 L^{\prime}$ kagom\'{e} 
layers within the periodic boundary (Fig.~\ref{3DPAF_2DPAF}). 
We used the Metropolis single spin-flip updates \cite{LandauBinder15} 
and the exchange Monte-Carlo method \cite{Hukushima96}. 
The CMC simulation software \cite{kadowaki_CMC_OH_2018} is based on 
an example of a Heisenberg model 
distributed by the ALPS project \cite{Bauer11,Albuquerque2007}. 
We note that the parameter set ($\delta, q$) had 
the substantial experimental uncertainty in Ref.~\cite{Takatsu2016prl}, 
which is shown by the elongated region enclosed by the dotted line 
in Fig.~1(a) of Ref.~\cite{Takatsu2016prl}. 
This uncertainty was concluded, 
because CMC simulations with small $\delta \ne 0$ show very similar results 
to those with $\delta = 0$ by adjusting the parameter $q$ \cite{Takatsu2016prl}. 

\section{Order parameters}
Long range orders of magnetic dipole and electric quadrupole moments 
expressed by pseudospin LRO 
$(\langle \sigma_{ \bm{r} }^x \rangle , \langle \sigma_{ \bm{r} }^y \rangle , \langle \sigma_{ \bm{r} }^z \rangle )$ 
were discussed using a classical mean-field analysis 
in zero field \cite{Onoda11}. 
It was shown that the PAF ordering has 
the highest mean-field critical temperature $T_{\text{c}}$ with 
degeneracy lines along [111] directions \cite{Onoda11}, 
more specifically, pseudospin LRO of non-zero $\langle \sigma_{ \bm{r} }^x \rangle$ 
and $\langle \sigma_{ \bm{r} }^y \rangle$ 
with modulation wavevectors $\bm{k}= (h, h, h)$ ($|h| \le \tfrac{1}{2}$). 
We summarize details of these classical mean-field LROs in the appendix. 
In addition, it was suggested \cite{Onoda11} that 
orders with the wavevector $\bm{k}=0$ can be selected 
from the infinitely degenerate mean-field PAF orders 
by an energetic \cite{PalmerChalker2000} or an order-by-disorder mechanism. 

The mean-field PAF order \cite{Onoda11} with a wavevector $\bm{k}= (h, h, h)$ 
is expressed by a pseudospin LRO 
\begin{equation}
\langle \bm{\sigma}_{ \bm{t}_n + \bm{d}_i } \rangle \propto \bm{v}_{i}^{\text{2D}} e^{i \bm{k} \cdot (\bm{t}_n + \bm{d}_i) } 
\label{PAF}
\end{equation}
with
\begin{equation}
\bm{v}_{i}^{\text{2D}} = 
\begin{cases} 
\bm{0} & (i=0) \\
\tfrac{\sqrt{3}}{2} \bm{x}_i  + \tfrac{1}{2} \bm{y}_i & (i=1) \\
-\tfrac{\sqrt{3}}{2} \bm{x}_i  + \tfrac{1}{2} \bm{y}_i & (i=2) \\
- \bm{y}_i & (i=3) \; ,
\end{cases}
\label{2DPAFk}
\end{equation}
where $\bm{x}_i$ and $\bm{y}_i$ stand for local axes 
at a crystallographic site $\bm{d}_i$ 
in the unit cell (Table~\ref{local_axis}), 
and $\bm{t}_n$ is an FCC translation vector. 
We note that these mean-field PAF orders have the zero amplitude on triangular 
lattice layers ($i=0$ sites in Fig.~\ref{3DPAF_2DPAF}), 
which implies that the PAF order is essentially 2D LRO 
on each kagom\'{e} lattice layer (appendix).

\subsection{Order parameter under zero magnetic field}
\begin{figure}
\centering
\includegraphics[width=7.0cm]{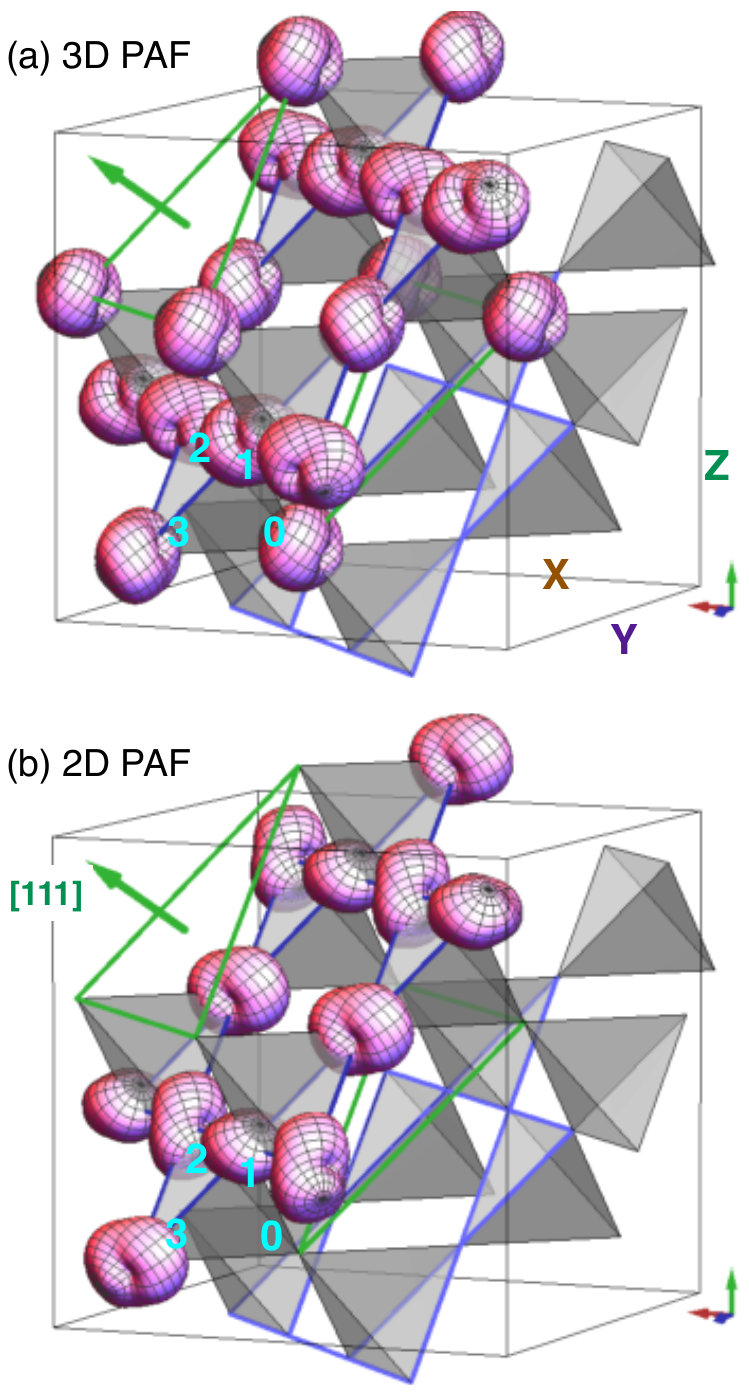}
\caption{ 
(a) 3D PAF [Eqs.~(\ref{3DPAF}) and (\ref{3DPAF_0})] and 
(b) 2D PAF [Eqs.~(\ref{2DPAF}) and (\ref{2DPAFk})] 
electric quadrupole orders are 
schematically illustrated by deformation of the $f$-electron 
change density from that of the paramagnetic phase \cite{Takatsu2016prl,Kadowaki15}.
}
\label{3DPAF_2DPAF}
\end{figure}
The one-fold degeneracy of the mean-field PAF order with a wavevector 
$\bm{k}= (h, h, h)$ ($h>0$) is increased to three-fold 
in the limit of $h \rightarrow 0$. 
These three pseudospin LRO structures 
with $\bm{k}=0$ are expressed by (appendix) 
\begin{equation}
\langle \bm{\sigma}_{ \bm{t}_n + \bm{d}_i } \rangle \propto \bm{v}_{i}^{(j)} \; ,
\label{3DPAF}
\end{equation}
where $j=0,1,2$ with 
\begin{equation}
\bm{v}_{i}^{(0)}=
\begin{cases} 
\bm{y}_i & (i=1,2) \\
-\bm{y}_i & (i=0,3) \; ,
\end{cases}
\label{3DPAF_0}
\end{equation}
\begin{equation}
\bm{v}_{i}^{(1)}=
\begin{cases} 
 \tfrac{\sqrt{3}}{2} \bm{x}_i - \tfrac{1}{2} \bm{y}_i & (i=1,3) \\
-\tfrac{\sqrt{3}}{2} \bm{x}_i + \tfrac{1}{2} \bm{y}_i & (i=0,2) \; ,
\end{cases}
\label{3DPAF_1}
\end{equation}
and 
\begin{equation}
\bm{v}_{i}^{(2)}= 
\begin{cases} 
 \tfrac{\sqrt{3}}{2} \bm{x}_i + \tfrac{1}{2} \bm{y}_i & (i=0,1) \\
-\tfrac{\sqrt{3}}{2} \bm{x}_i - \tfrac{1}{2} \bm{y}_i & (i=2,3) \; .
\end{cases}
\label{3DPAF_2}
\end{equation}

Under zero field, these 3D PAF orders can 
be stabilized energetically or by an order-by-disorder mechanism \cite{Onoda11}, 
which will be shown by CMC simulations. 
Their order parameters may be decomposed into 
\begin{equation}
m^{(j)}=\frac{\sum_{n,i} \bm{\sigma}_{ \bm{t}_n + \bm{d}_i } \cdot \bm{v}_{i}^{(j)} } {\sum_{n,i} 1} \; , 
\label{3DPAF_3op}
\end{equation}
where the summation runs over all sites $\bm{t}_n + \bm{d}_i$. 
In the limit of $T \rightarrow 0$, 
$(\langle m^{(0)} \rangle,\langle m^{(1)} \rangle,\langle m^{(2)} \rangle)$ becomes $(\pm 1,0,0)$, $(0,\pm 1,0)$, or $(0,0,\pm 1)$. 
In CMC simulations we measure the average of 
\begin{equation}
m_{\text{3DPAF}} = \sqrt{ [m^{(0)}]^2 + [m^{(1)}]^2 + [m^{(2)}]^2 } \; , 
\label{3DPAF_op}
\end{equation}
which represents the amplitude of the 3D PAF ordering. 

In Fig.~\ref{3DPAF_2DPAF}(a) we schematically illustrate 
the electric quadrupole order expressed by the pseudospin structure 
[Eqs.~(\ref{3DPAF}) and (\ref{3DPAF_0})]. 
We note that this ``3D PAF'' state is 
expressed by the ``$T_{2g}$'' state in Fig.~2(a) of Ref.~\cite{Rau_Gingras2018}, 
where a different notation is used: 
$J_{zz} = 4 J_{\text{nn,eff}}$, 
$J_{\pm}/J_{zz} = - \delta /2 $, 
and $J_{\pm \pm}/J_{zz} = q/2$; 
the local $\bm{x}_i$ and $\bm{y}_i$ 
are rotated by 120 degrees from our definition. 

\subsection{Order parameter under [111] magnetic field}
By taking a linear combination of Eq.~(\ref{PAF}) with various wavevectors $\bm{k}= (h, h, h)$ 
one can construct a 2D PAF pseudospin LRO 
which is non-zero only on an $\ell$-th kagom\'{e} lattice layer ($\ell=1,2, \cdots$) 
\begin{equation}
\langle \bm{\sigma}_{ \bm{t}_n + \bm{d}_i } \rangle \propto 
\bm{v}_{i}^{\text{2D}} \delta_{\ell,\hat{\bm{k}} \cdot (\bm{t}_n + \bm{d}_i) } \; ,
\label{2DPAF}
\end{equation}
where $\hat{\bm{k}}$ is a vector parallel to the [111] direction 
such that $\hat{\bm{k}} \cdot (\bm{t}_n + \bm{d}_i) = 1,2, \cdots$ on 
the kagom\'{e} layers. 
In Fig.~\ref{3DPAF_2DPAF}(b) we schematically illustrate 
the electric quadrupole order 
expressed by the pseudospin structure Eq.~(\ref{2DPAF}). 

Since mean fields on the triangular layers ($i=0$ sites) 
vanish for the 2D PAF order, 
magnetic dipole moments on the triangular layers, 
$\langle \sigma_{\bm{t}_n + \bm{d}_0}^z \rangle \bm{z}_0$, 
can be easily induced by applying [111] magnetic field. 
When this magnetized state is stabilized against 
the 3D PAF state by low [111] magnetic fields, 
one can expect that the system behaves as 
a 2D PAF state on each kagom\'{e} layer, 
which is decoupled by field-induced ferromagnetic triangular layers. 

Since $\bm{v}_{i}^{\text{2D}}$ 
[Eq.~(\ref{2DPAFk})] in 
Eq.~(\ref{2DPAF}) is expressed by 
$ \bm{v}_{i}^{\text{2D}} = \tfrac{1}{2} \left[ \bm{v}_{i}^{(0)} + \bm{v}_{i}^{(1)} + \bm{v}_{i}^{(2)} \right] $, 
we can define an order parameter of the 2D PAF order on a kagom\'{e} layer as 
\begin{equation}
m_{\text{2DPAF}} = \tfrac{2}{3} \left(  m^{(0) \prime} + m^{(1) \prime} + m^{(2) \prime} \right) 
\label{2DPAF_op}
\end{equation}
with 
\begin{equation}
m^{(j) \prime}=\frac{\sum_{n,i} \bm{\sigma}_{ \bm{t}_n + \bm{d}_i } \cdot \bm{v}_{i}^{(j)}}{\sum_{n,i} 1} \; , 
\label{2DPAF_3op}
\end{equation}
where the summation runs over sites 
on a single kagom\'{e} layer and an adjacent triangular layer. 
Under low [111] fields they become 
$\langle m^{(0) \prime} \rangle=\langle m^{(1) \prime} \rangle=\langle m^{(2) \prime} \rangle \simeq \pm \tfrac{1}{2}$ 
and $\langle m_{\text{2DPAF}} \rangle \simeq \pm 1$ at low temperatures. 
We will show that $m_{\text{2DPAF}}$ 
is the order parameter under low [111] fields by CMC simulations. 

\section{Results of CMC simulations}

\subsection{Zero magnetic field}
\begin{figure}
\centering
\includegraphics[width=8.7cm]{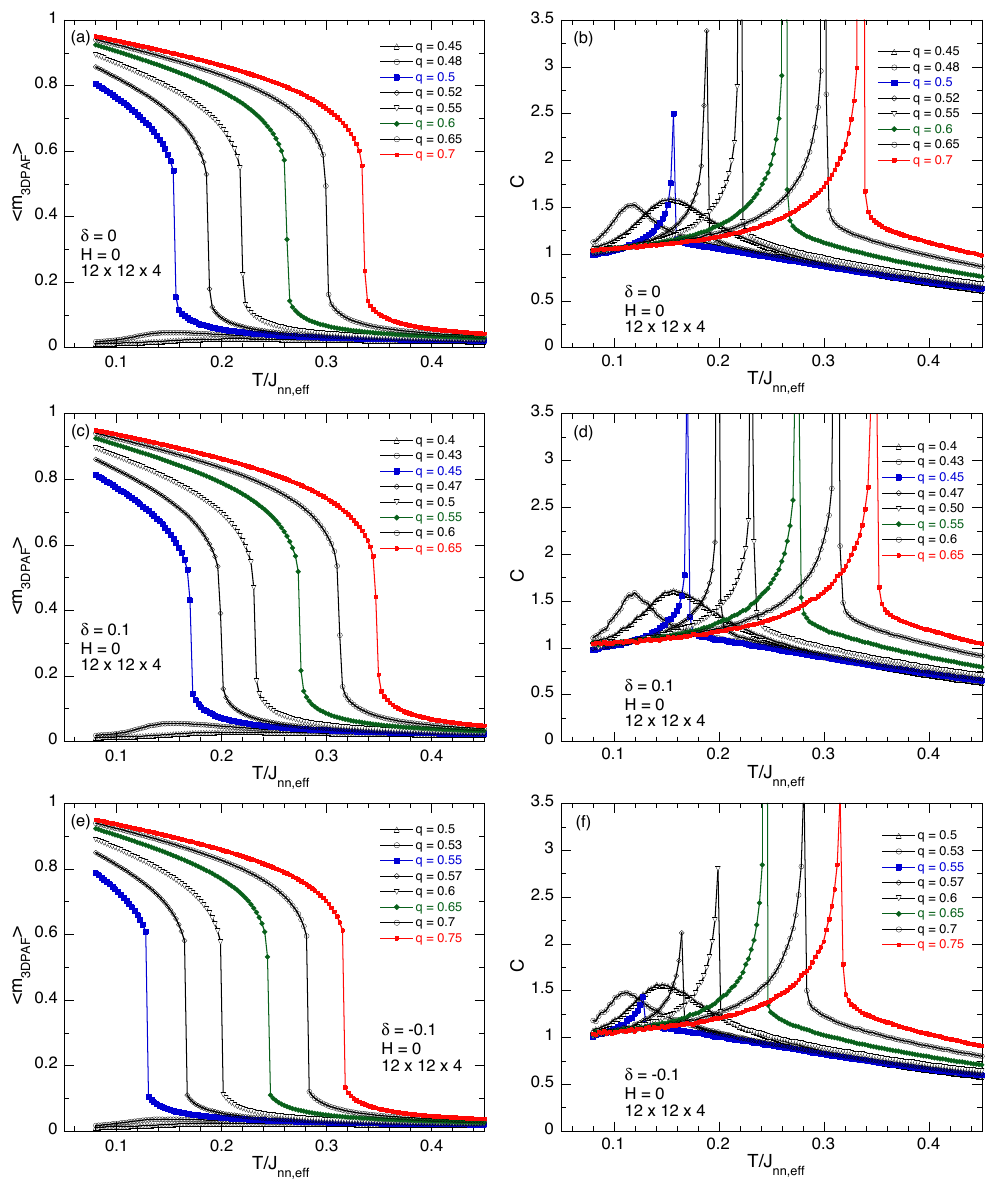}
\caption{ 
Temperature dependence of 
3D PAF order parameter $\langle m_{\text{3DPAF}} \rangle$ 
and specific heat $C$ under zero field 
calculated by CMC simulations for various $q$ values. 
Shown in (a,b) are results with $\delta = 0$; (c,d) and (e,f) are results 
with $\delta = 0.1$, and $-0.1$, respectively. 
}
\label{m3DPAF_C}
\end{figure}
Under zero magnetic field, it was shown that 
the classical ground state of the model 
for 
small $\delta$ 
changes from the classical spin ice state 
($q < q_{\text{c}}=(1-\delta)/2 $) 
to the PAF state ($q > q_{\text{c}}$) \cite{Onoda11}. 
We performed CMC simulations using several 
parameter sets of the effective Hamiltonian to 
clarify whether the energetic or the order-by-disorder selection 
mechanism stabilizes the 3D PAF order. 
The simulations were performed 
with a lattice size of $L=12$ and $L^{\prime}=4$ 
($12 \times 12 \times 4$).
In Fig.~\ref{m3DPAF_C} we plot 
the 3D-PAF order parameter $\langle m_{\text{3DPAF}} \rangle$ 
and 
the specific heat $C=(<E^2>-<E>^2)/(N T^2)$, where $E$ is the internal energy, 
as a function of temperature for 
$\delta=0, \pm 0.1$ 
and various $q$ values under zero field. 
One can see from Fig.~\ref{m3DPAF_C}(a) that 
$\langle m_{\text{3DPAF}} \rangle$ discontinuously increases 
below a critical temperature $T_{\text{c}}$ for 
$q \ge q_{\text{c}}$. 
This implies that the phase transition is first order and 
that the $\bm{k}=0$ order (3D PAF) occurs as expected. 
At the transition temperatures the specific heat [Fig.~\ref{m3DPAF_C}(b)] shows 
very sharp peaks. 
The CMC simulations with non-zero 
$\delta=0.1$ [Figs.~\ref{m3DPAF_C}(c) and (d)] 
and $\delta=-0.1$ [Figs.~\ref{m3DPAF_C}(e) and (f)] 
show parallel results with those of $\delta=0$. 
This confirms previous CMC simulations \cite{Takatsu2016prl} 
and is consistent with a mean-field result (appendix) 
that small $\delta$ only changes 
$T_{\text{c}}$ [the largest eigenvalue Eq.~(\ref{largest_eigenvalue})] 
as $T_{\text{c}}(q,\delta) = T_{\text{c}}(q,\delta=0) [1 + \delta /(2q)]$, 
without affecting eigenvectors 
Eqs.~(\ref{eigenvector2D}) and (\ref{eigenvector3D}). 

\begin{figure}
\centering
\includegraphics[width=7.5cm]{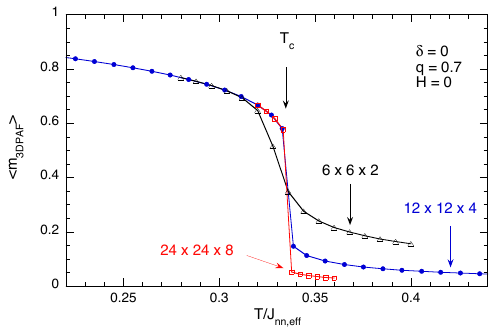}
\caption{ 
Size dependence of 3D PAF order parameter $\langle m_{\text{3DPAF}} \rangle$ 
as a function of temperature.}
\label{uu3D_size}
\end{figure}
%
\begin{figure}
\centering
\includegraphics[width=7.5cm]{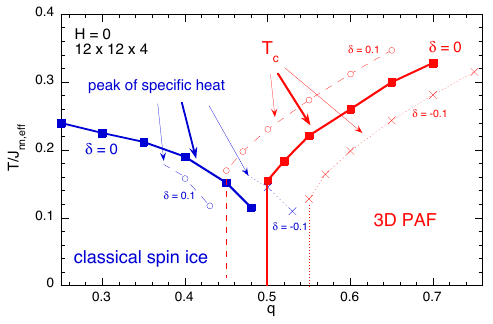}
\caption{ 
$T$-$q$ phase diagram determined by CMC simulations 
shown in Fig.~\ref{m3DPAF_C}. 
Red and blue thick lines are $T_{\text{c}}$ and 
broad peak of specific heat, respectively, obtained by simulations with $\delta = 0$. 
Dashed and dotted thin lines are those with $\delta = \pm 0.1$. 
}
\label{T-q_PhaseDiagram}
\end{figure}
Further CMC simulations with $(\delta,q) = (0,0.7)$ 
were performed to study size dependence of the 3D PAF order parameter. 
These results are shown in Fig.~\ref{uu3D_size}, 
which obviously demonstrates that the phase transition is first order. 
In Fig.~\ref{T-q_PhaseDiagram} three curves of $T_{\text{c}}$ 
are plotted as a function of $q$ for 
$\delta = -0.1$, $0.0$, and $0.1$. 
It discontinuously decreases to $T_{\text{c}}=0$ at the critical value 
$q_{\text{c}} = -0.45$, $0.5$, and $0.55$ for $\delta = -0.1$, $0.0$, and $0.1$, respectively. 
This agrees with the first-order nature of the quantum phase 
transition, which was investigated by a quantum treatment \cite{Lee12}. 
In the range $q < q_{\text{c}}$ the specific heat 
shows only a broad peak at about $T/J_{\text{nn,eff}} \sim 0.2$, 
which can be interpreted as the behavior of the classical spin ice 
model \cite{Onoda11}. 
We note that this peak temperature is significantly lower 
(about $1/4$) than that of the quantum MC simulation 
of the same model with parameters 
$q=0$ and $\delta \ne 0$ \cite{Kato15}. 
This implies that the temperature scale 
of the present CMC simulations is considerably reduced. 
Thereby one has to take account of this fact 
when comparing the CMC simulations 
with experimental data. 

\subsection{Under [111] magnetic field}
%
\begin{figure}
\centering
\includegraphics[width=7.5cm]{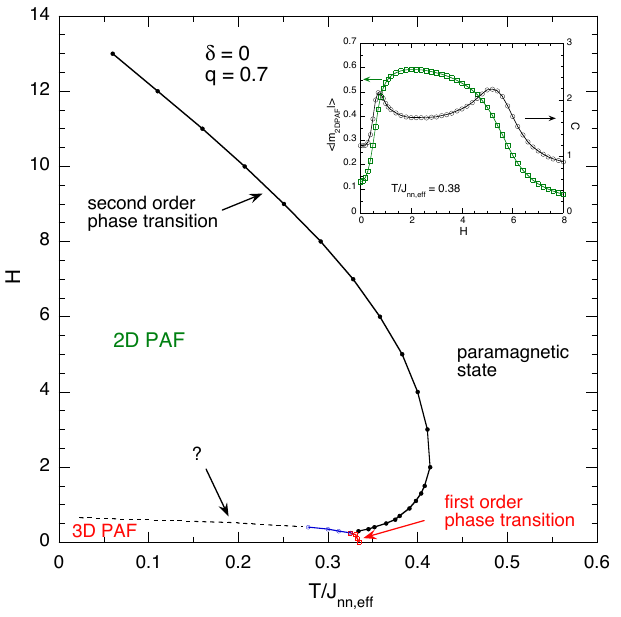}
\caption{ 
$H$-$T$ phase diagram under [111] magnetic field. 
There are the paramagnetic state and two LRO states 
of electric quadrupole moments denoted by 3D PAF and 2D PAF. 
Inset shows $H$ dependence of specific heat $C$ and 
2D PAF order parameter $\langle | m_{\text{2DPAF}} | \rangle$ 
at $T/J_{\text{nn,eff}} = 0.38$, 
which are calculated by simulations with lattice size $12 \times 12 \times 4$.
}
\label{HT_phase}
\end{figure}
To study finite-temperature phase-transitions under [111] magnetic fields, 
we performed CMC simulations 
with a parameter set $(\delta,q)=(0,0.7)$ 
under various fields $H$. 
Figure~\ref{HT_phase} shows an approximate $H$-$T$ phase diagram 
obtained from peaks of the specific heat and 
jumps of the order parameter $\langle m_{\text{3DPAF}} \rangle$, 
which are calculated by simulations with lattice sizes 
$12 \times 12 \times 4$ and/or $6 \times 6 \times 2$. 
From the high-temperature paramagnetic phase 
the system undergoes a phase transition to one of the 
two quadrupole ordered phases denoted by 3D PAF and 2D PAF, 
which will be discussed later. 

These 3D and 2D PAF phases are separated by a phase transition line, 
a crossover line, or multiple phase transitions 
(the dashed curve in Fig.~\ref{HT_phase}). 
These three possibilities could not be clarified by the present simulation 
techniques, because the single-spin-flip simulations suffer 
from a freezing problem at low temperatures. 
We note that the boundary line between 3D PAF and 2D PAF states 
depicted by the dashed curve in Fig.~\ref{HT_phase} 
corresponds to the low-field kink of the $M$-$H$ curve 
shown in Fig.~5(b) of Ref.~\cite{Takatsu2016prl}. 
Simulated $M(H,T)$ data suggest that there may 
be intermediate magnetization plateau states 
between zero field and the low-field kink. 

\begin{figure} 
\centering
\includegraphics[width=7.5cm]{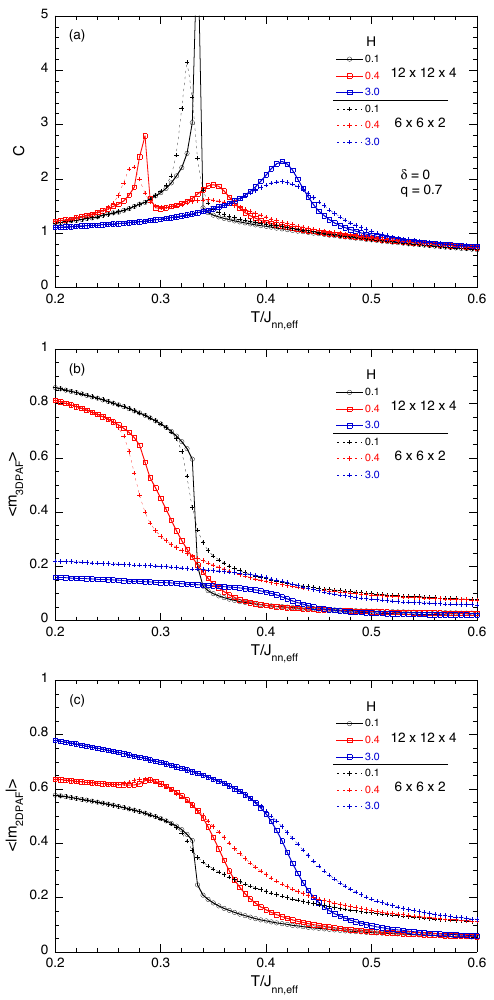}
\caption{ 
Temperature and size dependence of 
(a) specific heat $C$, 
(b) 3D PAF order parameter $\langle m_{\text{3DPAF}} \rangle$, and 
(c) 2D PAF order parameter $\langle | m_{\text{2DPAF}} | \rangle$
calculated by CMC simulations 
under three typical [111] fields $H=0.1$, 0.4, and 3. }
\label{C_uu3D_uu2D_h111}
\end{figure}
Figure~\ref{C_uu3D_uu2D_h111} shows 
temperature dependence of the specific heat $C$, 
the 3D-PAF order parameter $\langle m_{\text{3DPAF}} \rangle$, 
and the 2D-PAF order parameter $\langle | m_{\text{2DPAF}} | \rangle$ 
under three typical magnetic fields: 
$H=0.1$, 0.4, and 3. 
At the low field $H=0.1$ 
it is evident that the system shows 
the same first-order phase transition as zero field, 
and that LRO is the 3D PAF order. 
On the other hand, at the high field $H=3$, 
the size dependence of $C(T)$ and $\langle | m_{\text{2DPAF}} | \rangle(T)$ 
[Figs.~\ref{C_uu3D_uu2D_h111}(a) and \ref{C_uu3D_uu2D_h111}(c)] 
show typical behaviors of a second-order phase-transition. 
These indicate that $\langle | m_{\text{2DPAF}} | \rangle$ is the order parameter 
of the second-order phase-transition, 
in agreement with the initial expectation. 
At the intermediate field $H = 0.4$ 
the temperature dependence of 
the specific heat [Fig.~\ref{C_uu3D_uu2D_h111}(a)] implies that 
two successive phase transitions occur. 
At the higher $T_{\text{c}1}/J_{\text{nn,eff}} \simeq 0.35$,
$C(T)$ and $\langle | m_{\text{2DPAF}} | \rangle(T)$ 
[Figs.~\ref{C_uu3D_uu2D_h111}(a) and \ref{C_uu3D_uu2D_h111}(c)] 
show that 
the phase transition is the same kind as that for $H=3$. 
On the other hand, characteristics of the lower 
$T_{\text{c}2}/J_{\text{nn,eff}} \simeq 0.28$ 
are less clear owing to the freezing problem. 
The simulated $C(T)$, $\langle m_{\text{3DPAF}} \rangle(T)$, 
and $\langle | m_{\text{2DPAF}} | \rangle(T)$ 
(Fig.~\ref{C_uu3D_uu2D_h111})
suggest that $T_{\text{c}2}$ is 
a continuous phase transition between 2D PAF and 3D PAF states, 
which could not be further investigated using the present techniques. 
In addition to the constant $H$ plots (Fig.~\ref{C_uu3D_uu2D_h111}), 
magnetic field dependence of $C$ and $\langle | m_{\text{2DPAF}} | \rangle$ 
with constant $T=0.38 J_{\text{nn,eff}}$ 
are shown in the inset of Fig.~\ref{HT_phase}. 
At this temperature reentrant phase transitions occur 
at lower and upper critical fields, 
$H_{\text{c1}} \simeq 0.7$ and $H_{\text{c2}} \simeq 5.2$. 

\begin{figure} 
\centering
\includegraphics[width=7.5cm]{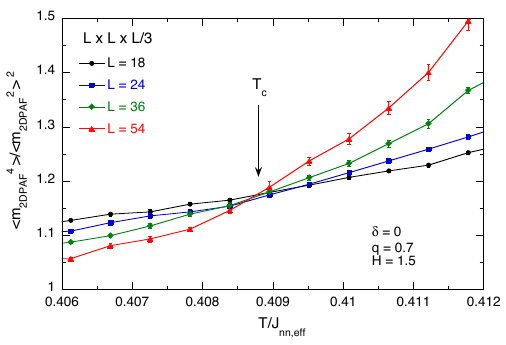}
\caption{ 
Temperature dependence of the Binder cumulant 
$\left< m_{\text{2DPAF}}^4 \right>/ \left< m_{\text{2DPAF}}^2 \right>^2$ 
close to $T_{\text{c}}$ 
for lattice sizes $L=18,24,36$, and $54$ under [111] field $H = 1.5$.
}
\label{BC_cross}
\end{figure}
Since the 2D PAF order breaks a $Z_2$ symmetry of $m_{\text{2DPAF}}$, 
one can naturally expect that its second-order phase-transition at $T_{\text{c}}$ 
belongs to the universality class of the 2D Ising model. 
To confirm this universality 
we performed standard finite-size scaling 
analyses \cite{LandauBinder15} 
on CMC simulation data taken 
under a typical [111] field $H=1.5$. 
These simulations were carried out 
on clusters with lattice sizes $L \times L \times (L/3)$ 
with $L=18,24,36$, and $54$. 
Figure~\ref{BC_cross} shows the Binder cumulant 
$U_4 = \left< m_{\text{2DPAF}}^4 \right>/ \left< m_{\text{2DPAF}}^2 \right>^2$ 
as a function of temperature. 
These curves with different lattice sizes cross at 
a single point, which enables us to 
determine the critical temperature $T_{\text{c}}/J_{\text{nn,eff}}=0.4088(2)$. 

\begin{figure} 
\centering
\includegraphics[width=7.5cm]{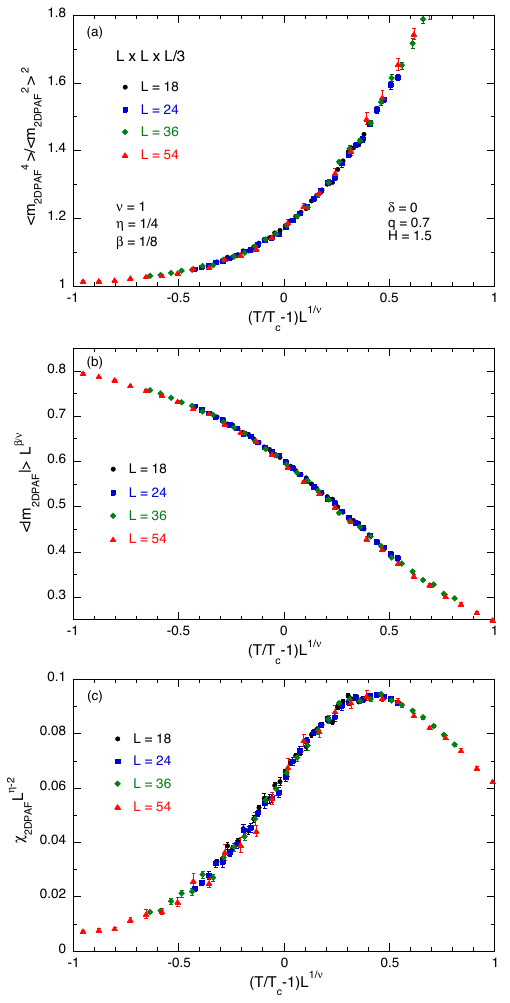}
\caption{ 
Finite size scaling of 
(a) the Binder cumulant $\left< m_{\text{2DPAF}}^4 \right>/ \left< m_{\text{2DPAF}}^2 \right>^2$, 
(b) 2D PAF order parameter $\langle | m_{\text{2DPAF}} | \rangle$, and 
(c) 2D PAF susceptibility $\chi_{\text{2DPAF}}$.}
\label{FS_scaling_plots}
\end{figure}
The theory of the finite-size scaling indicates that 
the Binder cumulant, 
the order parameter $\left< |m_{\text{2DPAF}}| \right>$, 
and the susceptibility 
$\chi_{\text{2DPAF}} = N_{\text{2D}} \left( \left< m_{\text{2DPAF}}^2 \right> - \left< |m_{\text{2DPAF}}| \right>^2 \right)/T$ 
show the scaling forms 
\begin{align}
U_4 &= f( L^{1/\nu} (T-T_{\text{c}})/T_{\text{c}} ) \; , \nonumber  \\
\left< |m_{\text{2DPAF}}|  \right> &= L^{-\beta/\nu} g( L^{1/\nu} (T-T_{\text{c}})/T_{\text{c}} ) \; , \\
\chi_{\text{2DPAF}} &= L^{2-\eta} h( L^{1/\nu} (T-T_{\text{c}})/T_{\text{c}} ) \; , \nonumber 
\label{FSscaling}
\end{align}
where $f$, $g$, and $h$ are universal functions \cite{LandauBinder15}. 
In Fig.~\ref{FS_scaling_plots} we show these finite-size scaling plots 
using the exact critical exponents $\nu=1$, $\beta=1/8$, and $\eta=1/4$ 
for the 2D Ising model. 
These figures show excellent data collapse, 
which proves the finite-size scaling relations of the 2D Ising model. 
Therefore we conclude that the second-order phase-transition 
of the 2D PAF state 
belongs to the 2D Ising universality class. 

\begin{figure} 
\centering
\includegraphics[width=7.5cm]{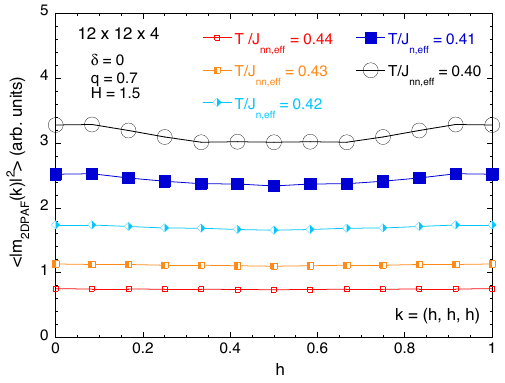}
\caption{ 
Wavevector dependence of $\langle |m_{\text{2DPAF}}(\bm{k})| \rangle^2$ 
along [111] direction above and below $T_{\text{c}}$ computed by CMC simulations 
with lattice size $12 \times 12 \times 4$. 
Size of symbol represents estimated error of data. 
}
\label{2D_Bragg}
\end{figure}

To complement the argument of the 2D Ising universality class 
we calculated squares of the Fourier transform of 
$m_{\text{2DPAF}}$ [Eq.~(\ref{2DPAF_op})], 
which is defined on each $\ell$-th kagom\'{e} lattice layer, 
with wavevectors $\bm{k}=(h,h,h)$ ($0 \le h \le 1$) 
\begin{equation}
|m_{\text{2DPAF}}(\bm{k})|^2 = \left| \sum_{\ell} \left[ m_{\text{2DPAF}} \right]_{\ell} \, e^{i \bm{k} \cdot \bm{r}} \right|^2 , 
\label{Bragg_int}
\end{equation}
where $\bm{r}$ is a lattice position on the $\ell$-th kagom\'{e} lattice layer. 
If $m_{\text{2DPAF}}$ has really 2D character, 
simulated averages of $|m_{\text{2DPAF}}(\bm{k})|^2$ do not depend on $h$. 
In terms of a scattering experiment (assuming that the quadrupole moment would be visible), 
$\langle |m_{\text{2DPAF}}(\bm{k})|^2 \rangle$ is constant 
between two $\Gamma$ points $\bm{k}= (0,0,0)$ and $(1,1,1)$. 
In Fig.~\ref{2D_Bragg} we show CMC averages $\langle |m_{\text{2DPAF}}(\bm{k})|^2 \rangle$ 
close to $T_{\text{c}}$, which were computed with a lattice size $12 \times 12 \times 4$. 
These curves show independence of $h$ and thereby the two dimensionality 
of the order parameter. 
We note that the freezing problem of the present CMC techniques 
prohibited us from performing simulations with larger system sizes 
and from obtaining the averages at low temperatures ($T \ll T_{\text{c}}$). 
This difficulty is seen as the large error estimation 
of the low-temperature data ($T \le T_{\text{c}}$) shown in Fig.~\ref{2D_Bragg}. 
Despite this large error, 
we also note that one may see slight wavevector dependence 
for the curve at $T=0.40 J_{\text{nn,eff}} < T_{\text{c}}$. 
This may suggest that the 2D PAF order is weakly modulated 
along the [111] direction at low temperatures. 
%

\section{Discussion}
In previous investigations \cite{Takatsu2016prl,Takatsu16} 
we showed that the simple pseudospin-$\frac{1}{2}$ Hamiltonian 
described by Eq.~(\ref{H_effective}) 
qualitatively and semi-quantitatively 
accounts for most of the experimental observations of the TTO sample 
with $T_{\text{c}}>0$ by selecting the appropriate model parameters. 
The agreement between experiments and theories 
was surprisingly better than our initial expectation. 
This means that the model Hamiltonian essentially explains 
the experimentally observed properties of TTO. 
Although there remain problems of oversimplifications caused by 
the classical approximations for the quantum model 
and by neglecting effects of higher-energy CF states \cite{Rau_Gingras2018} 
and Jahn-Teller effects due to 
the phonon mechanism \cite{Bonville11}. 

We would like to make a few comments 
on the the present CMC simulation results 
in relation to experimental observations. 
A first comment is on the natural question: 
how does the off-stoichiometry parameter 
of Tb$_{2+x}$Ti$_{2-x}$O$_{7+y}$, $x$ (and/or $y$), 
function as the tuning parameter between 
QSL and quadrupolar states? 
Our experiments using both poly- and single-crystalline samples 
showed that $x_{\text{c}} \simeq -0.0025$ is 
the quantum critical point \cite{Taniguchi13,Wakita16}. 
They also showed that by approaching to $x_{\text{c}}$ 
from the quadrupolar side $x > x_{\text{c}}$, 
the large specific-heat peak observed in 
$C(T)$ data (e.g. Fig.~4(a) in Ref.~\cite{Takatsu2016prl}) 
abruptly becomes smaller peaks 
as shown in Fig.~2 of Ref.~\cite{Taniguchi13} 
and Fig.~4(a) of Ref.~\cite{Wakita16}. 
By assuming that the change of $x$ is equivalent to that of $q$, 
the experimental behavior of $C(T)$ is approximately reproduced 
by the simulated $C(T)$ shown in Fig.~\ref{m3DPAF_C}(b). 
Therefore an answer to the question may be that 
$x$ tunes the ratio of the magnitude of 
the quadrupole interaction to that of the magnetic interaction.

\begin{figure}
\centering
\includegraphics[width=7.5cm]{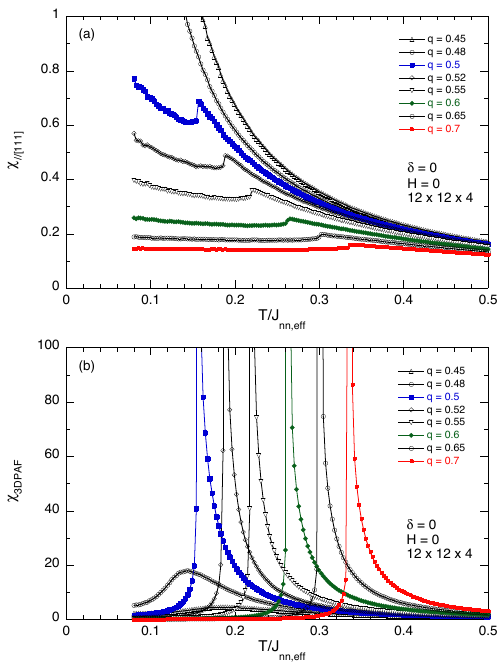}
\caption{ 
Temperature dependence of 
(a) magnetic susceptibility parallel to [111] direction $\chi_{\parallel [111]}$ 
and (b) susceptibility of $m_{\text{3DPAF}}$ under zero field 
calculated by CMC simulations. 
}
\label{chim3DPAF_chi111}
\end{figure}

A second comment is on susceptibilities under zero field. 
We calculated the magnetic susceptibility 
$\chi_{\parallel [111]} = N \left( \langle m_{\parallel [111]}^2 \rangle - \langle |m_{\parallel [111]}|\rangle^2 \right) /T$ 
using the same parameter sets as those of Fig.~\ref{m3DPAF_C}(a). 
These results are shown in Fig.~\ref{chim3DPAF_chi111}(a). 
The curve with $q=0.55$ 
bears resemblance to the experimental data 
of the TTO sample with $T_{\text{c}} = 0.53$ K 
(Fig.~2(a) of Ref.~\cite{Takatsu2016prl}). 
If we take account of the reduction of the temperature scale 
for the CMC simulation 
the resemblance becomes more striking. 
This also can justify the interpretation of TTO 
using the model Hamiltonian and the CMC simulation. 
We also calculated the electric quadrupole susceptibility 
corresponding to the 3D PAF order 
$\chi_{m_{\text{3DPAF}}} = N \left( \langle m_{\text{3DPAF}}^2 \rangle - \langle |m_{\text{3DPAF}}| \rangle^2 \right) /T$. 
Temperature dependence of this quadrupole susceptibility is 
shown in Fig.~\ref{chim3DPAF_chi111}(b). 
The large increase of $\chi_{m_{\text{3DPAF}}}$ close to $T_{\text{c}}$ 
can be measured by ultrasonic experiments of TTO, 
for example, extending measurements of Ref.~\cite{Nakanishi2011} down to 0.3 K. 

A third comment is on 
the first-order nature of the zero-field phase-transition 
of the CMC simulations. 
This does not agree with experimental $C(T)$, 
which shows a second-order behavior \cite{Takatsu2016prl}. 
In addition, the second-order phase-transition under [111] field 
seems to be somewhat smeared out 
for the the experimental data (Fig.~4(a,b) of Ref.~\cite{Takatsu2016prl}) 
compared to the CMC simulations. 
These disagreements remain to be explained, e.g.,  
by adding a higher-order term in the Hamiltonian \cite{Zhitomirsky2012}, 
by a disorder effect \cite{Imry75}, 
or possibly by a quantum effect. 

\section{Conclusions}
We have studied phase transitions of pyrochlore magnets 
with non-Kramers ions under [111] magnetic field 
represented by the effective pseudospin-$\frac{1}{2}$ Hamiltonian \cite{Onoda11} 
from a viewpoint of relevance to electric quadrupolar states 
of Tb$_{2}$Ti$_{2}$O$_{7}$ \cite{Takatsu2016prl}. 
Order parameters and finite-temperature 
phase-transitions of this frustrated model system 
are investigated using classical Monte-Carlo simulations. 
In zero field, the model undergoes a first-order phase-transition 
from the paramagnetic state to a 3D quadrupolar state 
with an antiparallel arrangement of pseudospins. 
This 3D order is selected energetically or by an order-by-disorder mechanism 
from degenerate $\bm{k}=(h,h,h)$ mean-field orders. 
Under [111] magnetic field this 3D state 
is transformed to a 2D quadrupolar state on each kagom\'{e} lattice,
which is separated by field-induced ferromagnetic triangular lattices. 
This 2D system undergoes a second-order phase-transition 
belonging to the 2D Ising universality class.
\\
\begin{acknowledgments}
We wish to thank S. Onoda and Y. Kato for useful discussions. 
This work was supported by JSPS KAKENHI grant numbers 25400345 and 26400336. 
\end{acknowledgments}

\appendix*
\section{\label{appendix_H_MF} Definitions of Hamiltonian and classical mean-field theory}
Detailed definitions of the Hamiltonian and 
pseudospin orders within a classical mean-field theory 
are summarized in this section. 
The CF ground state doublet of TTO \cite{Kadowaki15} can be written by 
\begin{equation}
| \pm 1 \rangle_{\text{D}} 
= A | \mp 4 \rangle \pm B | \mp 1 \rangle + C | \pm 2 \rangle \mp D | \pm 5 \rangle \; ,
\label{GD}
\end{equation}
where $| m \rangle$ stands for the $| J=6, m \rangle$ state within 
a $JLS$-multiplet \cite{Jensen91}. 
Using CF parameters of Ref.~\cite{Mirebeau07} 
the coefficients of Eq.~(\ref{GD}) are 
$A=0.9581$, $B=0.1284$, $C=0.1210$, $D=0.2256$. 
Magnetic-dipole and electric-quadrupole moment operators \cite{Kusunose08} 
within $| \pm 1 \rangle_{\text{D}} $ 
are proportional to the Pauli matrices $\sigma^{\alpha}$ ($\alpha = x,y,z$)
and the unit matrix \cite{Kadowaki15}: magnetic moment operators 
\begin{align}
J_x &= J_y = 0,\nonumber\\
J_z &= - (4A^2+B^2-2C^2-5D^2) \sigma^z \; ,
\label{mag}
\end{align}
and quadrupole moment operators 
\begin{align}
\tfrac{1}{2}[ 3 J_z^2 -J(J+1)] &= 
3A^2 - \tfrac{39}{2} B^2 - 15 C^2 + \tfrac{33}{2} D^2 , \nonumber\\
\tfrac{\sqrt{3}}{2}[ J_x^2 - J_y^2] &= 
\left( -\tfrac{21 \sqrt{3}}{2} B^2 + 9 \sqrt{10} AC \right) \sigma^x , \nonumber\\
\tfrac{\sqrt{3}}{2}[ J_x J_y + J_y J_x ] &= 
- \left( -\tfrac{21 \sqrt{3}}{2} B^2 + 9 \sqrt{10} AC \right) \sigma^y , \nonumber\\
\tfrac{\sqrt{3}}{2}[ J_z J_x + J_x J_z ] &= 
- \left( 3 \sqrt{30} BC + 9 \sqrt{\tfrac{33}{2}} AD \right) \sigma^x , \nonumber\\
\tfrac{\sqrt{3}}{2}[ J_y J_z + J_z J_y ] &= 
- \left( 3 \sqrt{30} BC + 9 \sqrt{\tfrac{33}{2}} AD \right) \sigma^y  \; .
\label{quad}
\end{align}

The operators $\sigma_{\bm{r}}^{\alpha}$ of Eq.~(\ref{H_effective}) 
act on $| \pm 1 \rangle_{\text{D}}$ 
at each pyrochlore lattice site $\bm{r}=\bm{t}_{n}+\bm{d}_{i}$, 
where $\bm{t}_n$ is an FCC translation vector and 
$\bm{d}_i$ ($i=0,1,2,3$) are four crystallographic sites in the unit cell. 
Coordinates of these sites $\bm{d}_i$ and their local axes 
$\bm{x}_i$, $\bm{y}_i$, and $\bm{z}_i$ 
are listed in Table~\ref{local_axis}. 
The phases $\phi_{\bm{r},\bm{r}^{\prime}}$ of Eq.~(\ref{H_effective}) are 
$\phi_{\bm{t}_n + \bm{d}_i, \bm{t}_{n^{\prime}} + \bm{d}_{i^{\prime}}} = 0$, 
$- 2 \pi /3$, and $2 \pi /3$ 
for site pairs of $(i, i^{\prime}) = (0,3), (1,2)$, 
$(i, i^{\prime}) = (0,1), (2,3)$, 
and $(i, i^{\prime}) = (0,2), (1,3)$, respectively, 
where the notation of Ref.~\cite{Onoda11} is used. 
\begin{table}
\caption{\label{local_axis}
Coordinates of four crystallographic sites $\bm{d}_{i}$ and their 
local axes $\bm{x}_{i}$, $\bm{y}_{i}$, and $\bm{z}_{i}$ \cite{Kadowaki15}. 
These coordinates are defined using (global) cubic XYZ axes shown in Fig.~\ref{3DPAF_2DPAF}(a).
The four sites $\bm{d}_{i}$ are illustrated by vertices with light blue numbers ($i=0,1,2,3$) of a tetrahedron in Fig.~\ref{3DPAF_2DPAF}(a). 
}
\begin{ruledtabular}
\begin{tabular}{ccccc}
$i$ & $\bm{d}_{i}$ & $\bm{x}_{i}$ & $\bm{y}_{i}$ & $\bm{z}_{i}$ \\ \hline
0 & $\tfrac{1}{4}(0,0,0)$ & $\tfrac{1}{\sqrt{6}}(1,1,-2)$ & $\tfrac{1}{\sqrt{2}}(-1,1,0)$ & $\tfrac{1}{\sqrt{3}}(1,1,1)$  \\
1 & $\tfrac{1}{4}(0,1,1)$ & $\tfrac{1}{\sqrt{6}}(1,-1,2)$ & $\tfrac{1}{\sqrt{2}}(-1,-1,0)$ & $\tfrac{1}{\sqrt{3}}(1,-1,-1)$  \\
2 & $\tfrac{1}{4}(1,0,1)$ & $\tfrac{1}{\sqrt{6}}(-1,1,2)$ & $\tfrac{1}{\sqrt{2}}(1,1,0)$ & $\tfrac{1}{\sqrt{3}}(-1,1,-1)$  \\
3 & $\tfrac{1}{4}(1,1,0)$ & $\tfrac{1}{\sqrt{6}}(-1,-1,-2)$ & $\tfrac{1}{\sqrt{2}}(1,-1,0)$ & $\tfrac{1}{\sqrt{3}}(-1,-1,1)$  \\
\end{tabular}
\end{ruledtabular}
\end{table}

Possible pseudospin LROs of Eq.~(\ref{H_effective}) under zero magnetic field 
were discussed in Ref.~\cite{Onoda11}. 
We summarize a few results of the classical mean-field theory \cite{Onoda11} 
to facilitate gaining insight of order parameters for the PAF phase 
(Fig.~7 in Ref.~\cite{Onoda11}; $q>q_{\text{c}}$). 
The effective Hamiltonian of Eq.~(\ref{H_effective}) under zero magnetic field 
can be expressed using the Fourier transform as 
\begin{equation}
\mathcal{H} \propto - J_{\text{nn,eff}} \sum_{\bm{k} , i, i^{\prime}, \alpha, \beta} 
 \sigma_{\bm{k},i}^{\alpha} J_{i, \alpha; i^{\prime}, \beta}(\bm{k}) \sigma_{\bm{k},i^{\prime}}^{\beta} 
 \; ,
\label{FT}
\end{equation}
where the summation runs over wavevectors $\bm{k}$ in the first Brillouin zone, 
$i, i^{\prime} = 0,1,2,3$ and $\alpha, \beta = x,y,z$, and 
$\sigma_{\bm{t}_n + \bm{d}_i}^{\alpha} = \sum_{\bm{k}} \sigma_{\bm{k},i}^{\alpha} e^{i \bm{k} \cdot (\bm{t}_n + \bm{d}_i)}$. 
The matrix $J_{i, \alpha; i^{\prime}, \beta}(\bm{k})$ stands for the Fourier transform 
of the superexchange coupling constants $J_{n, i, \alpha; n^{\prime}, i^{\prime}, \beta}$ 
between $\sigma_{\bm{t}_n + \bm{d}_i}^{\alpha}$ and $\sigma_{\bm{t}_{n^{\prime}} + \bm{d}_{i^{\prime}}}^{\beta}$: 
\begin{equation}
J_{i, \alpha; i^{\prime}, \beta}(\bm{k}) = 
\sum_{n} J_{n, i, \alpha; n^{\prime}, i^{\prime}, \beta} 
e^{i \bm{k} \cdot [ (\bm{t}_{n}+\bm{d}_{i}) - (\bm{t}_{n^{\prime}}+\bm{d}_{i^{\prime}}) ]} \;.
\label{J_Ftransform}
\end{equation}
The critical temperature $T_{\text{c}}$ and pseudospin LRO 
are obtained by the largest eigenvalue ($\propto T_{\text{c}}$) 
and corresponding eigenvectors of $J_{i, \alpha; i^{\prime}, \beta}(\bm{k})$. 

The largest eigenvalue of $J_{i, \alpha; i^{\prime}, \beta}(\bm{k})$ is degenerate 
on four symmetry-equivalent lines  
$\bm{k}= (h,\pm h,h)$ and $(h,h,\pm h)$, where $|h| \le \tfrac{1}{2}$ \cite{Onoda11}. 
On a degeneracy line $\bm{k}= (h,h,h)$, 
the $12 \times 12$ matrix $J_{i, \alpha; i^{\prime}, \beta}(\bm{k})$ consists of 
magnetic $4 \times 4$ and quadrupolar $8 \times 8$ blocks: the magnetic submatrix 
\begin{align}
& J_{i, z; i^{\prime}, z}(\bm{k}=(h,h,h)) = - J_{\text{nn,eff}} \nonumber\\
& \times 
\begin{pmatrix} 
0 & \cos (\pi h) & \cos (\pi h) & \cos (\pi h) \\ 
\cos (\pi h) & 0 & 1 & 1 \\
\cos (\pi h) & 1 & 0 & 1 \\
\cos (\pi h) & 1 & 1 & 0 \\
\end{pmatrix}
 \; ,
\label{Jzz}
\end{align}
which acts on a vector 
$(\sigma_{\bm{k},0}^{z},\sigma_{\bm{k},1}^{z},\sigma_{\bm{k},2}^{z},\sigma_{\bm{k},2}^{z})^{\text{T}}$, 
and the quadrupolar submatrix 
\begin{align}
& J_{i, \alpha; i^{\prime}, \beta}(\bm{k}=(h,h,h)) = - J_{\text{nn,eff}} \nonumber\\
& \times 
\begin{pmatrix} 
0 & \cos (\pi h) M_1 & \cos (\pi h) M_2 & \cos (\pi h) M_3 \\ 
\cos (\pi h) M_1 & 0 & M_3 & M_2 \\
\cos (\pi h) M_2 & M_3 & 0 & M_1 \\
\cos (\pi h) M_3 & M_2 & M_1 & 0 \\
\end{pmatrix}
 \; ,
\label{Jxy}
\end{align}
which acts on a vector 
$(\sigma_{\bm{k},0}^{x},\sigma_{\bm{k},0}^{y},\sigma_{\bm{k},1}^{x},\sigma_{\bm{k},1}^{y},\sigma_{\bm{k},2}^{x},\sigma_{\bm{k},2}^{y},\sigma_{\bm{k},3}^{x},\sigma_{\bm{k},3}^{y})^{\text{T}}$. 
In Eq.~(\ref{Jxy}) $M_i$ ($i=1,2,3$) stand for $2 \times 2$ matrices 
$M_1 = 
\begin{pmatrix} 
\delta - \tfrac{1}{2}q & -\tfrac{\sqrt{3}}{2}q   \\
-\tfrac{\sqrt{3}}{2}q   & \delta + \tfrac{1}{2}q \\
\end{pmatrix}$ \, , 
$M_2 = 
\begin{pmatrix} 
\delta - \tfrac{1}{2}q &  \tfrac{\sqrt{3}}{2}q   \\
 \tfrac{\sqrt{3}}{2}q   & \delta + \tfrac{1}{2}q \\
 \end{pmatrix}$ \, , 
and 
$M_3 = 
\begin{pmatrix} 
\delta + q & 0          \\
0          & \delta - q \\
\end{pmatrix}$. 
One can show that the largest eigenvalue of $J_{i, \alpha; i^{\prime}, \beta}(\bm{k})$ 
is that of Eq.~(\ref{Jxy}), which is exactly 
\begin{equation}
J_{\text{nn,eff}}(2q + \delta) 
\label{largest_eigenvalue}
\end{equation}
for small $\delta$ (PAF phase). 

One can also show that 
the degeneracy of the largest eigenvalue 
is one and three fold for $|h|>0$ and $h=0$, respectively, 
and that the corresponding eigenvectors, 
which depend on neither $q$ nor $\delta$, 
are given by 
\begin{equation}
\begin{pmatrix}
\sigma_{\bm{k},0}^{x} \\
\sigma_{\bm{k},0}^{y} \\
\sigma_{\bm{k},1}^{x} \\
\sigma_{\bm{k},1}^{y} \\
\sigma_{\bm{k},2}^{x} \\
\sigma_{\bm{k},2}^{y} \\
\sigma_{\bm{k},3}^{x} \\
\sigma_{\bm{k},3}^{y} \\
\end{pmatrix}
=
\begin{pmatrix}
0 \\
0 \\
\tfrac{\sqrt{3}}{2} \\
\tfrac{1}{2} \\
-\tfrac{\sqrt{3}}{2} \\
\tfrac{1}{2} \\
0 \\
-1 \\
\end{pmatrix}
\label{eigenvector2D}
\end{equation}
[Eq.~(\ref{2DPAFk})] for $|h|>0$ 
and by 
\begin{equation}
\begin{pmatrix}
\sigma_{\bm{0},0}^{x} \\
\sigma_{\bm{0},0}^{y} \\
\sigma_{\bm{0},1}^{x} \\
\sigma_{\bm{0},1}^{y} \\
\sigma_{\bm{0},2}^{x} \\
\sigma_{\bm{0},2}^{y} \\
\sigma_{\bm{0},3}^{x} \\
\sigma_{\bm{0},3}^{y} \\
\end{pmatrix}
=
\begin{pmatrix}
0 \\
-1 \\
0 \\
1 \\
0 \\
1 \\
0 \\
-1 \\
\end{pmatrix}
\, ,
\begin{pmatrix}
-\tfrac{\sqrt{3}}{2} \\
\tfrac{1}{2} \\
\tfrac{\sqrt{3}}{2} \\
-\tfrac{1}{2} \\
-\tfrac{\sqrt{3}}{2} \\
\tfrac{1}{2} \\
\tfrac{\sqrt{3}}{2} \\
-\tfrac{1}{2} \\
\end{pmatrix}
\, ,
\begin{pmatrix}
\tfrac{\sqrt{3}}{2} \\
\tfrac{1}{2} \\
\tfrac{\sqrt{3}}{2} \\
\tfrac{1}{2} \\
-\tfrac{\sqrt{3}}{2} \\
-\tfrac{1}{2} \\
-\tfrac{\sqrt{3}}{2} \\
-\tfrac{1}{2} \\
\end{pmatrix}
\label{eigenvector3D}
\end{equation}
[Eqs.~(\ref{3DPAF_0}), (\ref{3DPAF_1}), (\ref{3DPAF_2})]
for $h=0$. 
Therefore, it is very likely that  
pseudospin LROs of Eq.~(\ref{H_effective}) just below $T_{\text{c}}$ 
under zero magnetic field 
are either the mean-field PAF order [Eq.~(\ref{eigenvector2D})] 
or the 3D PAF order [Eq.~(\ref{eigenvector3D})]. 
Although it is not obvious which PAF order is selected, 
one can expect that at sufficiently low temperatures 
an energetic or an order-by-disorder mechanism 
stabilizes the 3D PAF order. 
We note that for the PAF order [Eq.~(\ref{eigenvector2D})] 
the mean field at the triangular lattice site ($\bm{d}_{i=0}$) vanishes, 
which implies that the PAF order is essentially 2D LRO on each kagom\'{e} lattice layer. 

\bibliography{TTO_MC_HK}
\end{document}